\documentclass[conference,letter]{IEEEtran}
\IEEEoverridecommandlockouts
\usepackage{cite}
\usepackage{amsmath,amssymb,amsfonts}
\usepackage{algorithmic}
\usepackage{graphicx}
\usepackage{textcomp}
\usepackage{xcolor}
\usepackage{booktabs}
\usepackage{subcaption}
\usepackage{physics}

\makeatletter %
\newcommand{\linebreakand}{%
  \end{@IEEEauthorhalign}
  \hfill\mbox{}\par
  \mbox{}\hfill\begin{@IEEEauthorhalign}
}
\makeatother %

\begin{document}

\title{Metamorphic Testing with the Rashomon Set:\\Explanation Faithfulness in Machine Learning\thanks{This work is supported by the Norwegian Ministry of Education and Research (Kunnskapsdepartementet).}
}

\author{%
\IEEEauthorblockN{Helge Spieker}
\IEEEauthorblockA{\textit{Simula Research Laboratory}\\
Oslo, Norway \\
helge@simula.no}
\and
\IEEEauthorblockN{Jørn Eirik Betten}
\IEEEauthorblockA{\textit{Simula Research Laboratory}\\
Oslo, Norway \\
jorneirik@simula.no}
\and
\IEEEauthorblockN{Arnaud Gotlieb}
\IEEEauthorblockA{\textit{Simula Research Laboratory}\\
Oslo, Norway \\
arnaud@simula.no}%
}
\maketitle

\begin{abstract}
Multiple machine learning models can achieve near-equivalent predictive
performance on the same task, yet provide divergent
feature-based explanations.
This is called the Rashomon effect of (explainable) machine learning, and it raises
the question of which explanations, if any, are trustworthy.
We propose a framework based on metamorphic testing that assesses explanation faithfulness without requiring ground-truth labels by exploring attributed feature importance from post-hoc explanation methods.
Five metamorphic relations formalize expected consistency properties
between model behavior and feature attributions.
We apply this general framework to two tabular regression datasets and two post-hoc explainers (SHAP and LIME) to demonstrate the approach.
The framework offers a practical, model-agnostic tool for selecting accurate
models with reliable and trustworthy explanations.
\end{abstract}


\section{Introduction}

The Rashomon effect in machine learning describes the situation in which many models fit the training data equally well but tell different stories about which features matter~\cite{breiman2001}. 
When such models are paired with post-hoc explanation methods, stakeholders (end users, regulators, domain experts) may receive conflicting accounts of the same phenomenon depending on which model is selected, undermining trust and actionability~\cite{Muller2023}.

Most existing faithfulness metrics for explainable AI (XAI) require ground-truth attributions or synthetic data with known feature relevance~\cite{hooker2019}. 
These oracles are unavailable in real-world settings, leaving practitioners without a principled way to decide which explanations to trust when models disagree.

Investigating explanation faithfulness is difficult because there is generally a discrepancy between what dataset analysis techniques might indicate as the most prevalent and relevant features (in addition to the common presence of feature interactions) and what the model interprets as most relevant features for making a prediction.
Similarly, although there is sometimes a coupling between the numerical feature importance value and its impact on the prediction outcome, it is not strictly given or independent of other features, making the definition of precise test oracles challenging.

Metamorphic testing (MT)~\cite{Chen2018}, on the other hand, is particularly suited for test scenarios with an oracle problem~\cite{Barr2015}, that is, when it is difficult to specify the precise expected outcome, as described above.

In this work, we propose an approach towards metamorphic testing of explanation faithfulness in the context of the Rashomon effect in machine learning. Our proposed method assesses explanation faithfulness by measuring the consistency between a model's predictions and its explanations, without requiring ground truth. 
Here, we present our approach consisting of:
(1) five metamorphic relations that encode expected relationships
between attributions and model sensitivity, including both single-model and cross-model consistency; 
(2) five corresponding aggregate metrics that summarize test outcomes;
and (3) a test pipeline built on $\varepsilon$-Rashomon sets of well-performing models.
We demonstrate our approach through an evaluation on two tabular regression benchmarks, that is, California Housing (8 features) and Wine Quality (11 features), using SHAP and LIME as competing post-hoc explainers on heterogeneous candidate pools of 85 models each across six model families.
All metrics are compared against a random-attribution null baseline.
Empirical results reveal systematic differences in explanation faithfulness both within and across explainers, with LIME producing notably more homogeneous explanations across Rashomon-set members than SHAP.

The remainder of the paper is organized as follows: Section~\ref{sec:background} introduces the background and related work. Section~\ref{sec:method} introduces our method for MT with the Rashomon set, including the metamorphic relations and the overall testing pipeline. Sections~\ref{sec:experiments} and~\ref{sec:results} describe the experimental setup and the results, which are followed by an overall discussion in Section~\ref{sec:discussion}, before Section~\ref{sec:conclusion} concludes the paper.

\section{Background and Related Work}\label{sec:background}

\subsection{The Rashomon Effect}

The Rashomon effect, introduced by Breiman~\cite{breiman2001}, refers to the existence of multiple models that achieve similar predictive accuracy while implementing different functional forms. Formally, the $\varepsilon$-Rashomon set collects all models whose loss is within a relative tolerance $\varepsilon$ of the best observed loss. Its existence has significant implications for model selection, variable importance analysis, and the scientific conclusions drawn from data~\cite{semenova2022,marx2020,fisher_all_2019}.

Previous work has also investigated the existence of the Rashomon effect for Explainable AI (XAI)~\cite{Muller2023,Spieker2025}, showing that two models may provide different explanations for the same prediction, raising the question of explanation faithfulness: how reliable and trustworthy are explanations?

\subsection{Explanation Faithfulness}

Post-hoc explanation methods such as SHAP~\cite{lundberg2017} and LIME~\cite{ribeiro2016} aim to approximate a model's reasoning. A central question is whether such explanations are \emph{faithful}, i.e., whether they reflect the model's actual predictive behavior~\cite{adebayo2018,rudin2019}. Established faithfulness measures include insertion/deletion curves, sensitivity analysis, and axiom-based evaluations~\cite{hooker2019}. These approaches typically require ground-truth attributions, synthetic benchmarks, or single-model evaluation, limiting their applicability to real-world datasets and to settings where model multiplicity is explicit.

\subsection{Metamorphic Testing}

Metamorphic testing~\cite{segura2016,Chen2018} is an oracle-free software testing technique that formalizes expected input--output relationships as metamorphic relations (MRs). Prior work has applied metamorphic testing to machine learning systems~\cite{Xie2009,Dwarakanath2018,Dwarakanath2019,Spieker2020,Xiao22,xu2024evaluating,Spieker2024,Spieker2025a,duran2025metamorphic}, and their improvement~\cite{Xu2018,DBLP:journals/jss/XuTFBZC21}, focusing on the consistency of \emph{predictions} under input transformations, both on trained and untrained models.
A first, related approach of metamorphic testing to evaluate explainability has been presented by Fan et al.~\cite{Fan2022}. Their work focuses on the dedicated evaluation of object detection models and the influence of semantic changes, i.e., object insertion, object removal, and background manipulation, on the corresponding explanations.
In our work, we aim to broaden this perspective and operate on the direct level of feature importance, providing a complementary perspective.

\section{Method}\label{sec:method}

\subsection{Problem Formulation}

Let $\{f_1, \dots, f_k\}$ be a set of trained models with similar test loss (the $\varepsilon$-Rashomon set), and let $E$ be a post-hoc explainer that produces an attribution vector $a = E(f, x) \in \mathbb{R}^d$ for model $f$ at input $x \in \mathbb{R}^d$, with $a_i$ being the importance score assigned to feature $i$ and $|a_j|$ denoting its absolute value. 
Additionally, we denote an input perturbed in feature $i$ as $x_i' = x + \epsilon e_i$, where $e_i$ is a basis vector with all zeros except $e_i=1$, such that only feature $i$ is perturbed. 
Our goal is to assess whether each model's explanations faithfully reflect its predictive behavior, and whether explanations are consistent across models.
Our metamorphic relations (MRs) are therefore the expected changes in the predictive behavior under changes to the features in relation to their assessed importance.
The MR is violated if the expected change does not occur, indicating an unreliable explainer with an unfaithful model explanation.

\subsection{Metamorphic Relations and Metrics}

For each MR, we state its intuition, what a violation reveals, and the aggregate metric used to quantify it.

\textbf{MR1: Explanation Faithfulness.} If $|a_j| > |a_k|$ for features $j$ and $k$, then perturbing feature $j$ should change $f(x)$ more than perturbing feature $k$ by the same magnitude. A violation indicates that the attribution's ranking does not reflect the model's actual sensitivity ranking of features.

We report both a strict binary version (FS) and a continuous Spearman rank correlation between $|a|$ and $|\Delta f|$ across all features ($\text{FS}_{rank}$), which makes graded differences between explainers visible.

\emph{Faithfulness Score (FS):} 
This is the per-model fraction of test inputs where perturbing the top-attributed feature moves the prediction more than perturbing the bottom-attributed feature. 
$\text{FS}_{rank}$ is the continuous analogue: the Spearman rank correlation between $|a|$ and $|\Delta f|$ averaged over all test inputs.

\emph{Area over the Perturbation Curve (AOPC):} 
As an additional, comparative metric to the faithfulness score, we consider the metric in which the most or least relevant features are iteratively removed, and the change in prediction is measured.
AOPC removes features in rank order, that is, first the most-attributed (MoRF; most relevant first), then the least-attributed (LeRF; least relevant first), and then measures the cumulative drop in model output at each step~\cite{samek2016evaluating}. 
A faithful explainer should cause a larger drop when removing high-attribution features first.
So, $AOPC_{MoRF} - AOPC_{LeRF} > 0$ indicates that the attributions correctly identify what the model relies on. It is a decision-level alternative to the faithfulness score.

\textbf{MR2: Cross-Model Sensitivity Consistency.} 
If two models $f_i$ and $f_j$ agree on the most important feature, perturbing that feature should shift both predictions in the same direction.

A violation indicates that a shared explanation does not imply shared behavior.

\emph{Rashomon Explanation Agreement (REA):} Per-input fraction of
Rashomon-set model pairs that, when agreeing on a top feature, also agree on the direction of the prediction shift induced by the perturbation, averaged over test inputs.

\textbf{MR3: Explanation Divergence Implies Sensitivity Divergence.} If two models cite different top features, each model should be more sensitive to its own cited feature than the other model is. A violation indicates that divergent explanations do not reflect genuinely different model strategies.

\emph{Divergence Consistency (DC):} Per-input fraction of Rashomon-set model pairs for which \emph{disagreement} on the top feature is matched by each model being more sensitive to its own cited feature than to the other's; averaged over test inputs. 
To extract a fuller picture of the explanation behavior, we measure both $\text{DC}_\text{all}$ (all pairs, even those that do not disagree) and $\text{DC}_\text{app}$ (restricted to pairs where the top features actually disagree), together with the applicability rate. %

\textbf{MR4: Invariance Under Irrelevant Transformations.} Perturbing features with near-zero attribution should not meaningfully change predictions or explanations. 
A violation indicates that ``unimportant'' features covertly influence the model.

\emph{Explanation Fragility Index (EFI):} Per-model fraction of low-attribution (``irrelevant'') features whose perturbation nonetheless shifts the prediction or the explanation beyond a tolerance. Here, a lower value indicates greater alignment between the explanation and the model's behavior.

\textbf{MR5: Proportional Attribution Response.} The ratio of output shifts from perturbing the top two features should approximately equal the ratio of their attributions. A violation indicates that attribution magnitudes are not quantitatively calibrated, even if they are ordinally correct.

\emph{Proportionality Score (PS):} Per-model fraction of test inputs where the ratio of prediction shifts from perturbing the top two attributed features is within a tolerance $\gamma$ of the ratio of their attributions.

\subsection{Testing Pipeline}

The testing method proceeds in four stages. \emph{First}, we construct the $\varepsilon$-Rashomon set by selecting all trained candidates whose test error is within $\varepsilon_r$-relative tolerance of the best observed error.
\emph{Second}, each model is paired with a post-hoc explainer: we evaluate SHAP~\cite{lundberg2017} and LIME~\cite{ribeiro2016} independently on the same Rashomon set, enabling comparison of MR outcomes across explainers.
Attributions from both explainers are peak-normalized to $[0,1]$ to make the MR thresholds comparable.
\emph{Third}, we apply a feature-space perturbation (additive shift) of magnitude $\delta$ and evaluate the MRs on a subsample of $n_\text{test}$ test inputs.
\emph{Finally}, all five metrics with 95\% bootstrap confidence intervals (CIs) are reported, and results are compared against a random-attribution null baseline.

Note that our testing pipeline does not follow the traditional software testing setup, in which we have a clear binary pass/fail definition (even in the context of metamorphic testing), but we are additionally interested in effect sizes and effect differences between systems under test. The exact behavior of explainers depends strongly on the individual models, and from the existing literature, we know that there is variability in model behavior and explainability, which can lead to similar variability in the provided explanations. That means, a violation of the MR does not necessarily mean a failure of the explainer per se, but it is an effect that warrants further analysis and awareness.

\section{Experimental Setup}\label{sec:experiments}

\subsection{Datasets}

We evaluate on two tabular regression benchmarks.
\emph{California Housing}~\cite{pace1997} (20,640 samples, 8 features) is a well-studied dataset with non-trivial feature correlations with the target to predict median house values.
\emph{Wine Quality}~\cite{cortez2009} (1,599 samples, 11 physicochemical features) is a dataset on wines, with the aim to predict the sensory quality score of red wines.
In our experiments, all features are standardized, and we use a fixed 80/20 train/test split for all training runs.

\subsection{Candidate Pool and Model Families}

The candidate pool is identical for both datasets and consists of 85 models
drawn from six families:
(i) \emph{Linear} (Ridge, Lasso, ElasticNet);
(ii) \emph{PolyLinear} (degree-2/3 polynomial features plus Ridge/Lasso);
(iii) \emph{Kernel} (KernelRidge (RBF, polynomial) and SVR);
(iv) \emph{Tree} (RandomForest and ExtraTrees in standard and deep variants);
(v) \emph{Boosting} (GradientBoosting, deep GBR, HistGradientBoosting,
and regularised HistGB);
(vi) \emph{Neural Networks} (MLP with varying width, depth, and regularisation).
All stochastic models are trained with four random seeds.

\subsection{$\varepsilon$-Rashomon Set Selection}

\begin{figure}[t]
  \centering
  \includegraphics[width=0.9\columnwidth]{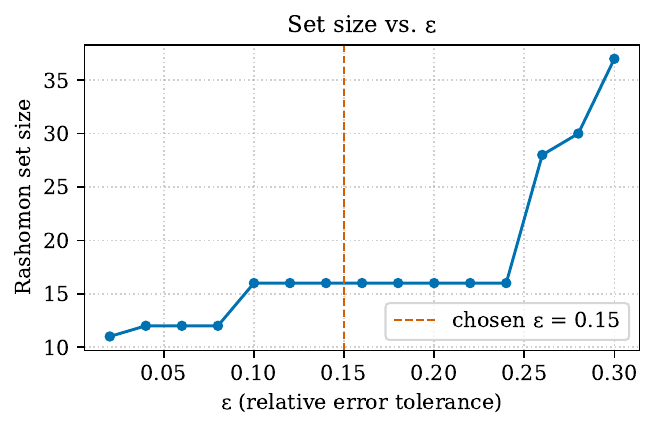}
  \caption{Rashomon set size vs.\ relative error tolerance $\varepsilon_r$
    on California Housing. The dashed line marks the chosen
    threshold $\varepsilon_r = 0.15$; the set stabilises at 16 models
    from $\varepsilon_r \approx 0.10$ onwards.}
  \label{fig:eps_sweep}
\end{figure}

We adopt $\varepsilon_r = 0.15$ for both datasets: the Rashomon set contains every model whose MSE lies within 15\% of the best observed MSE.
Figure~\ref{fig:eps_sweep} shows how set size varies with $\varepsilon_r$; we select $\varepsilon_r = 0.15$ as it is a stabilized midpoint in the sweep results. 

In California Housing, the best MSE is $0.200$, and the 15\%
threshold admits 16 models, all from the Boosting family (gradient-boosting variants outperform every other family by a large margin at this dataset size and feature scale).
On Wine Quality, the best MSE is $0.282$, and the threshold admits 27 models from two families: 16 Tree (RandomForest and ExtraTrees variants) and 11 Boosting models.
Again, most model families are not performing well enough to be included in the Rashomon set.

\section{Results}\label{sec:results}

\begin{figure*}[t]
  \centering
  \includegraphics[width=\textwidth]{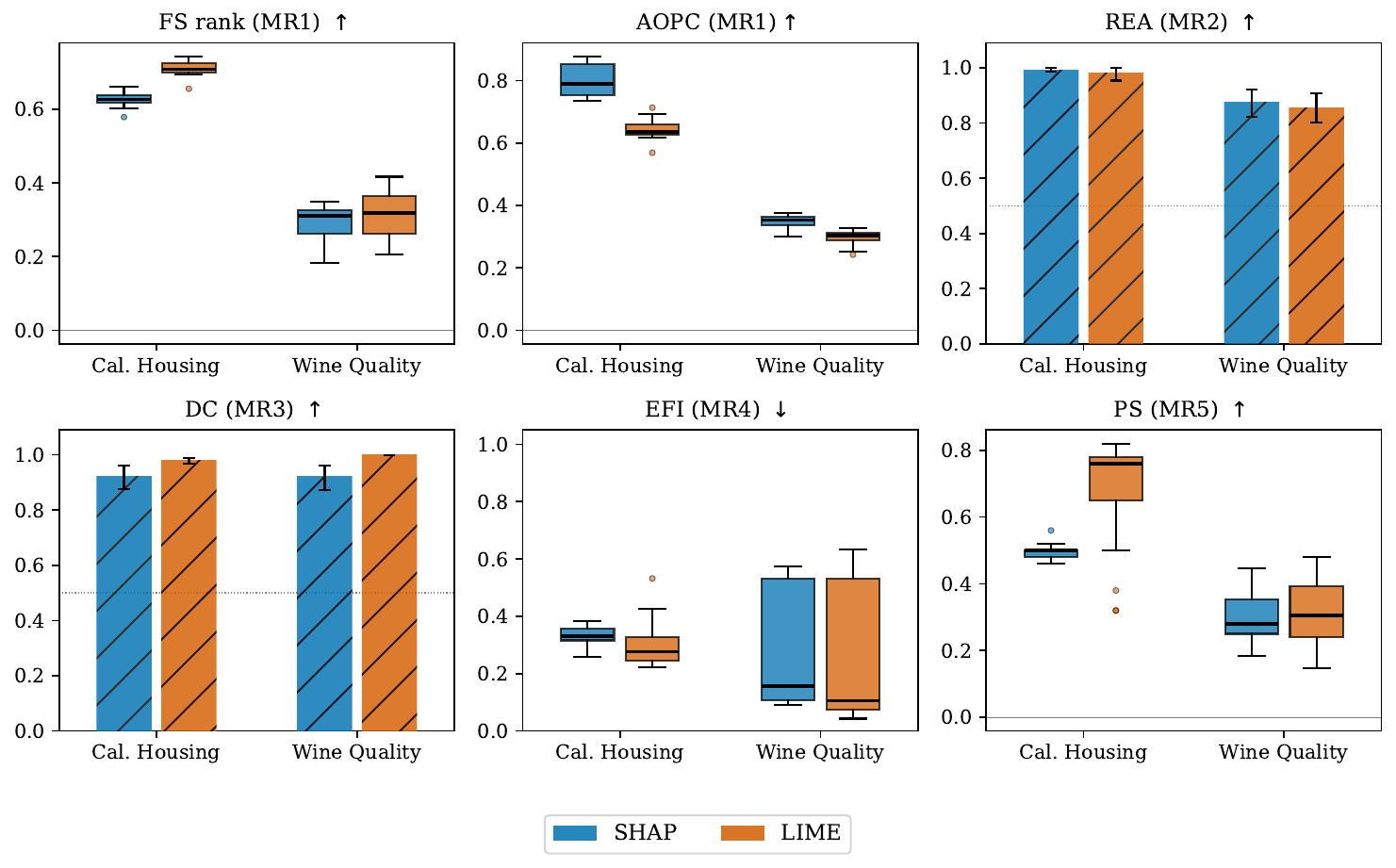}
  \caption{MR metric comparison across explainers and datasets.
    Rows: datasets. Columns: $\text{FS}_{rank}$ (MR1),
    AOPC (MR1), REA (MR2), DC (MR3), EFI (MR4), PS (MR5).
    Aggregate metrics (REA, DC; striped bars) show point estimates with 95\% bootstrap CIs.
    Distribution metrics ($\text{FS}_{rank}$, AOPC, EFI, PS) show individual
    Rashomon-set model values with mean line.}
  \label{fig:mr_comparison}
\end{figure*}

The main outcomes of the experiments are shown in Table~\ref{tab:violations}, which lists the percentage of MR violations observed over the entire test process. 
The table is organized by datasets, explainers, and MR.
Additionally, in Figure~\ref{fig:mr_comparison}, we summarize the overall results to compare the two explainers side-by-side for each MR.
In what follows, we discuss and interpret these results while providing additional perspectives and summaries of the overall outcomes.

\begin{table*}[t]
  \centering
  \caption{MR violation rates averaged over Rashomon-set models and test
    inputs. MR2 (REA) and MR3 (DC) are reported as the violation rate among
    applicable pairs only, with applicability rate in brackets; n/a denotes
    negligible applicability ($<0.2$\%). All $\text{FS}_{rank}$ values are far above
    the random-attribution null ($\rho \approx 0.00$); null EFI~$= 1.00$.}
  \label{tab:violations}
  \begin{tabular}{lllcccc}
    \toprule
    & & & \multicolumn{2}{c}{California Housing} & \multicolumn{2}{c}{Wine Quality} \\
    \cmidrule(lr){4-5}\cmidrule(lr){6-7}
    MR & Name & Metric & SHAP & LIME & SHAP & LIME \\
    \midrule
    MR1 & Explanation Faithfulness & Violation (1$-$FS) $\downarrow$ & 6\%  & 6\%  & 26\% & 20\% \\
        & & $\text{FS}_{rank}$ (Spearman $\rho$) $\uparrow$ & 0.626 & 0.711 & 0.290 & 0.313 \\
    \midrule
    MR2 & Cross-Model Sensitivity & Violation [app.\ rate] $\downarrow$ & 1\% [90\%] & 2\% [97\%] & 14\% [88\%] & 14\% [100\%] \\
    \midrule
    MR3 & Explanation Divergence & Violation [app.\ rate] $\downarrow$ & 76\% [10\%] & 72\% [3\%] & 66\% [12\%] & n/a [$<$1\%] \\
    \midrule
    MR4 & Invariance under Irrelevant Transformations & EFI (mean) $\downarrow$             & 0.334 & 0.303 & 0.281 & 0.260 \\
    \midrule
    MR5 & Proportional Attribution Response & Violation (1$-$PS) $\downarrow$     & 51\%  & 33\%  & 70\%  & 69\%  \\
    \bottomrule
  \end{tabular}
\end{table*}

\subsection{MR1: Explanation Faithfulness}

In terms of violations, MR1 (Faithfulness Score) shows mild binary violation on the California Housing set (6\%), but increases to 20--26\% on the more challenging Wine Quality task, where eleven correlated chemical features make it harder for the explainer to correctly rank sensitivity.

In the California Housing case, for SHAP, $\text{FS}_{rank}$ ranges from 0.58 to 0.66 (mean~$\approx$~0.62), and AOPC ranges from 0.73 to 0.88, confirming genuine per-model faithfulness.
For LIME, the $\text{FS}_{rank}$ distribution is similar but slightly more compressed.

In the Wine Quality case, under SHAP, $\text{FS}_{rank}$ is lower than on California Housing (range 0.20--0.35),
consistent with the harder attribution task on 11 correlated chemical
features. AOPC is similarly lower (0.27--0.36).

All $\text{FS}_{rank}$ values are far above the random-attribution null baseline ($\rho_\text{null} \approx 0.00$), confirming that the violations are not artifacts of the metric design.

\subsection{MR2: Cross-Model Sensitivity Consistency}

MR2 evaluates the sensitivity to feature perturbations between two models. It is measured by the Rashomon Explanation Agreement (REA).

It has a high applicability rate (88--100\%), of which a moderate number of applicable pairs violate the MR (circa 14\%). This result is similar for both LIME and SHAP, showing that neither captures the direction of change under feature perturbation entirely correct.
Still, it is low enough to indicate that indeed top features are correctly identified in most cases.

\subsection{MR3: Explanation Divergence}

MR3 measures the difference in sensitivity of the top features between model pairs that disagree on these top features, i.e., it focuses on the sensitivity arising from diverging importance.

For both datasets, we observe that LIME's explanation diversity is consistently lower than SHAP's.
DC applicability is 10--12\% for SHAP but near zero for LIME on both datasets. LIME's local linear surrogates tend to identify the same dominant feature across all models in the set, whereas SHAP's model-structure-aware values preserve variation across models.
For wine quality, this rate is essentially zero ($<0.1$\%), meaning all 27 Rashomon models cite the same top feature across all test inputs, despite belonging to two different families.
In consequence, $\text{DC}_\text{all}(\text{LIME}) = 1.00\ [1.00, 1.00]$ (see Figure~\ref{fig:mr_comparison}), but this reflects the absence of divergence rather than active validation.

This is a case where the Rashomon effect in Explainable Machine Learning~\cite{Muller2023} does not occur; the models indeed mirror the true feature importance. 
However, note that this is a comparatively simple dataset, with potentially little redundancy and data symmetries that could be (inadvertently) exploited by the models; a potential reason for the Rashomon effect in XAI for larger datasets~\cite{Spieker2025}.

\subsection{MR4: Invariance Under Irrelevant Transformations}

MR4 focuses on the lowest-ranking features, which should have the least influence on the output prediction.

For both datasets, EFI is relatively consistent, with SHAP slightly higher than LIME, and California Housing having a higher EFI than Wine Quality.
On average, 26--33\% of nominally irrelevant features still influence predictions or explanations beyond tolerance; this indicates that ``irrelevance'' as labeled by the explainer is often not irrelevance in the model's input--output mapping.
This does not necessarily mean that the explainer is incorrect: first, all features are actually important, and a ranking cannot reveal that; second, feature interactions might still involve low-ranking features (as measured over the entire dataset) for certain inputs and predictions. %

\subsection{MR5: Proportional Attribution Response}

For MR5, we observe more substantial differences, both between the two explainers and between the datasets. It is among the most frequently violated MRs, alongside MR3 (Explanation Divergence).
Especially for California Housing, the PS score is much higher, indicating a greater proportional response in the prediction to the feature change.
PS violation rates range from 33\% to 70\%; attribution magnitudes are poorly calibrated even with correct ordinal ranking (MR1).

\section{Discussion}\label{sec:discussion}

\subsection{Key Results}

A key finding is that LIME explanations are nearly identical across heterogeneous Rashomon-set members in both datasets (near-zero DC applicability), whereas SHAP retains measurable cross-model divergence.
When explanations diverge (under SHAP), they are rarely supported by genuine sensitivity differences (DC$_\text{app}$ near null), suggesting that explanation differences within Rashomon sets are often artifacts of explainer behavior rather than model behavior.
The proposed metrics (FS/$\text{FS}_{rank}$, REA, DC, EFI, PS) with bootstrap CIs, applicability rates, and null baselines provide actionable, statistically grounded criteria for selecting models whose explanations can be trusted.

At the same time, while our experiments show the principal applicability of our metamorphic testing methodology, the results raise questions for future work and additional experiments:
The low applicability rates for MR3 (Explanation Divergence) indicate that the Rashomon effect in explainable machine learning is only weak on these datasets, an uncommon result in the existing literature. 
A practitioner who observes low DC applicability (as seen with LIME) should not interpret this as ``the Rashomon effect is absent'': it means that the chosen explainer assigns the same top feature to all models regardless of their structural differences. Whether this reflects genuine feature dominance or explainer homogenization remains to be investigated.
To better understand this effect, an extended evaluation across additional datasets and model families is required, as well as to determine whether the substantial difference between SHAP and LIME is an experimental artifact or a fundamental difference (with potentially broad relevance).

\subsection{What Metamorphic Testing Reveals That Other Methods Do Not}

No ground-truth attributions are needed, making the framework applicable to any real-world dataset. It tests \emph{consistency} between model and explanation rather than \emph{correctness} against a synthetic oracle. The cross-model relations MR2 and MR3 are uniquely enabled by the Rashomon-set framing; single-model evaluation cannot detect the cross-model phenomena they capture. The DC applicability rate is a novel diagnostic: near-zero applicability does not necessarily indicate that explanations are aligned, but rather signals that the explainer may be suppressing genuine model diversity.

\subsection{Limitations}

Our experimental framework includes several methodological choices and limitations that provide avenues for deeper investigation and extensions in future work:

\paragraph{Perturbation and Sampling}
Perturbation-based testing risks generating out-of-distribution inputs, and our results are sensitive to the perturbation magnitude, which we set based on preliminary experiments rather than an exhaustive sensitivity study. 
To ensure tractability, we limit our test set to $n_{test}=50$ instances, mitigating sampling uncertainty through 95\% bootstrap confidence intervals.

\paragraph{Hyperparameters and Metrics} 
The threshold $\varepsilon_r=0.15$ is an empirical selection, though sensitivity to this parameter is detailed in Figure~\ref{fig:eps_sweep}. 
Furthermore, binary MR1 (FS) evaluates only extreme-ranked features. While $\text{FS}_{rank}$ covers the full feature space, its dependence on Spearman rank correlation makes it sensitive to ties. Similarly, MR5 selects only the top two features, as they are deemed most important, but the MR could be extended to any pair of features for further investigation, under acceptance of a higher computational budget.

\paragraph{Models and Modalities} The study focuses on standard scikit-learn algorithms applied to two tabular regression datasets. Future work should explore whether the observed faithfulness–fragility profiles generalize to more complex deep learning models (or even large language models) and to high-dimensional text or image data, following the work in \cite{Fan2022}.

\section{Conclusion}\label{sec:conclusion}

We have introduced a methodology for applying metamorphic testing in conjunction with the Rashomon set of well-performing models to evaluate the explanation faithfulness of machine learning explanation methods.
At the same time, the methodology enables model selection based on explanation quality rather than predictive accuracy alone.

The Rashomon effect poses a concrete risk for explanation-driven decision-making: equally accurate models can produce conflicting and unreliable explanations. 
Metamorphic testing provides a principled, oracle-free framework for testing explanation faithfulness by checking internal consistency between predictions and attributions.
Our empirical results on California Housing and Wine Quality, each evaluated under SHAP and LIME, show that explanation quality varies systematically across datasets, model families, and explainers.

Beyond this specific, exemplary results, we demonstrated the applicability of the approach, and provided the foundation for future work and investigations on the usage of metamorphic testing to understand the effects that stem from a multiplicity of good models, i.e., the Rashomon effect, and its connection to explainable AI.
Future work will need to analyze more complex data sources, including other modalities such as text or images, and other model architectures, particularly deep learning architectures.

\bibliographystyle{IEEEtran}
\bibliography{refs.bib}

\begin{thebibliography}{10}
\providecommand{\url}[1]{#1}
\csname url@samestyle\endcsname
\providecommand{\newblock}{\relax}
\providecommand{\bibinfo}[2]{#2}
\providecommand{\BIBentrySTDinterwordspacing}{\spaceskip=0pt\relax}
\providecommand{\BIBentryALTinterwordstretchfactor}{4}
\providecommand{\BIBentryALTinterwordspacing}{\spaceskip=\fontdimen2\font plus
\BIBentryALTinterwordstretchfactor\fontdimen3\font minus
  \fontdimen4\font\relax}
\providecommand{\BIBforeignlanguage}[2]{{%
\expandafter\ifx\csname l@#1\endcsname\relax
\typeout{** WARNING: IEEEtran.bst: No hyphenation pattern has been}%
\typeout{** loaded for the language `#1'. Using the pattern for}%
\typeout{** the default language instead.}%
\else
\language=\csname l@#1\endcsname
\fi
#2}}
\providecommand{\BIBdecl}{\relax}
\BIBdecl

\bibitem{breiman2001}
L.~Breiman, ``Statistical modeling: The two cultures (with comments and a
  rejoinder by the author),'' \emph{Statistical science}, 2001.

\bibitem{Muller2023}
S.~M{\"u}ller, V.~Toborek, K.~Beckh, M.~Jakobs, C.~Bauckhage, and P.~Welke,
  ``An {Empirical} {Evaluation} of the {Rashomon} {Effect} in {Explainable}
  {Machine} {Learning},'' in \emph{{ECML-PKDD}}, 2023.

\bibitem{hooker2019}
S.~Hooker, D.~Erhan, P.~Kindermans, and B.~Kim, ``A benchmark for
  interpretability methods in deep neural networks,'' in \emph{Advances in
  Neural Information Processing Systems 32 (NeurIPS)}, 2019, pp. 9734--9745.

\bibitem{Chen2018}
T.~Y. Chen, F.-C. Kuo, H.~Liu, P.-L. Poon, D.~Towey, T.~H. Tse, and Z.~Q. Zhou,
  ``Metamorphic {Testing}: {A} {Review} of {Challenges} and {Opportunities},''
  \emph{ACM Computing Surveys}, 2018.

\bibitem{Barr2015}
E.~T. Barr, M.~Harman, P.~McMinn, M.~Shahbaz, and S.~Yoo, ``The {Oracle}
  {Problem} in {Software} {Testing}: {A} {Survey},'' \emph{IEEE Transactions on
  Software Engineering}, vol.~41, no.~5, 2015.

\bibitem{semenova2022}
L.~Semenova, C.~Rudin, and R.~Parr, ``On the existence of simpler machine
  learning models,'' in \emph{{ACM} Conference on Fairness, Accountability, and
  Transparency ({FAccT})}, 2022.

\bibitem{marx2020}
C.~Marx, F.~Calmon, and B.~Ustun, ``Predictive multiplicity in
  classification,'' in \emph{Proceedings of the 37th International Conference
  on Machine Learning (ICML)}, 2020.

\bibitem{fisher_all_2019}
A.~Fisher, C.~Rudin, and F.~Dominici, ``All {Models} are {Wrong}, but {Many}
  are {Useful}: {Learning} a {Variable}’s {Importance} by {Studying} an
  {Entire} {Class} of {Prediction} {Models} {Simultaneously},'' \emph{{JMLR}},
  vol.~20, 2019.

\bibitem{Spieker2025}
H.~Spieker, J.~E. Betten, A.~Gotlieb, N.~Lazaar, and N.~Belmecheri, ``Rashomon
  in the streets: Explanation ambiguity in scene understanding,''
  \emph{Proceedings of the AAAI Symposium Series (ATRACC)}, no.~1, 2025.

\bibitem{lundberg2017}
S.~M. Lundberg and S.-I. Lee, ``A unified approach to interpreting model
  predictions,'' in \emph{Advances in Neural Information Processing Systems
  (NeurIPS)}, vol.~30, 2017.

\bibitem{ribeiro2016}
M.~T. Ribeiro, S.~Singh, and C.~Guestrin, ``"why should {I} trust you?":
  Explaining the predictions of any classifier,'' in \emph{{ACM} {SIGKDD}
  International Conference on Knowledge Discovery and Data Mining}, 2016.

\bibitem{adebayo2018}
J.~Adebayo, J.~Gilmer, M.~Muelly, I.~J. Goodfellow, M.~Hardt, and B.~Kim,
  ``Sanity checks for saliency maps,'' in \emph{Advances in Neural Information
  Processing Systems ({{NeurIPS}})}, 2018, pp. 9525--9536.

\bibitem{rudin2019}
C.~Rudin, ``Stop explaining black box machine learning models for high stakes
  decisions and use interpretable models instead,'' \emph{Nat. Mach. Intell.},
  vol.~1, no.~5, pp. 206--215, 2019.

\bibitem{segura2016}
S.~Segura, G.~Fraser, A.~B. S{\'{a}}nchez, and A.~R. Cort{\'{e}}s, ``A survey
  on metamorphic testing,'' \emph{{IEEE} Trans. Software Eng.}, 2016.

\bibitem{Xie2009}
X.~Xie, J.~Ho, C.~Murphy, G.~Kaiser, B.~Xu, and T.~Chen, ``Application of
  {{Metamorphic Testing}} to {{Supervised Classifiers}},'' in \emph{2009
  {{Ninth Int. Conf.}} on {{Quality Software}}}, vol.~33, 2009, pp. 135--144.

\bibitem{Dwarakanath2018}
A.~Dwarakanath, M.~Ahuja, S.~Sikand, R.~M. Rao, R.~P. J.~C. Bose, N.~Dubash,
  and S.~Podder, ``Identifying implementation bugs in machine learning based
  image classifiers using metamorphic testing,'' in \emph{{ACM} {SIGSOFT}
  {International} {Symposium} on {Software} {Testing} and {Analysis}
  ({ISSTA})}, 2018.

\bibitem{Dwarakanath2019}
A.~Dwarakanath, M.~Ahuja, S.~Podder, S.~Vinu, A.~Naskar, and M.~Koushik,
  ``Metamorphic {Testing} of a {Deep} {Learning} {Based} {Forecaster},'' in
  \emph{2019 {IEEE}/{ACM} 4th {International} {Workshop} on {Metamorphic}
  {Testing} ({MET})}, May 2019, pp. 40--47.

\bibitem{Spieker2020}
H.~Spieker and A.~Gotlieb, ``Adaptive metamorphic testing with contextual
  bandits,'' \emph{Journal of Systems and Software}, vol. 165, 2020.

\bibitem{Xiao22}
D.~Xiao, Z.~LIU, Y.~Yuan, Q.~Pang, and S.~Wang, ``Metamorphic testing of deep
  learning compilers,'' in \emph{Proceedings of the ACM on Measurement and
  Analysis of Computing Systems}, vol.~6, 2022.

\bibitem{xu2024evaluating}
Y.~Xu, Y.~Li, J.~Wang, and X.~Zhang, ``Evaluating terminology translation in
  machine translation systems via metamorphic testing,'' in \emph{Proceedings
  of the 39th IEEE/ACM International Conference on Automated Software
  Engineering}, 2024, pp. 758--769.

\bibitem{Spieker2024}
H.~Spieker, N.~Belmecheri, A.~Gotlieb, and N.~Lazaar, ``Evaluating {Human}
  {Trajectory} {Prediction} with {Metamorphic} {Testing},'' in \emph{{ACM}
  {International} {Workshop} on {Metamorphic} {Testing} ({MET})}, 2024.

\bibitem{Spieker2025a}
H.~Spieker, N.~Lazaar, A.~Gotlieb, and N.~Belmecheri, ``Metamorphic testing of
  multimodal human trajectory prediction,'' \emph{Information and Software
  Technology}, p. 107890, 2025.

\bibitem{duran2025metamorphic}
M.~Duran, T.~Laurent, E.~Rushe, and A.~Ventresque, ``Metamorphic testing for
  pose estimation systems,'' in \emph{2025 IEEE Conference on Software Testing,
  Verification and Validation (ICST)}, 2025.

\bibitem{Xu2018}
L.~Xu, D.~Towey, A.~P. French, S.~Benford, Z.~Q. Zhou, and T.~Y. Chen,
  ``Enhancing supervised classifications with metamorphic relations,''
  \emph{3rd International Workshop on Metamorphic Testing - MET '18}, 2018.

\bibitem{DBLP:journals/jss/XuTFBZC21}
------, ``Using metamorphic relations to verify and enhance artcode
  classification,'' \emph{J. Syst. Softw.}, vol. 182, p. 111060, 2021.

\bibitem{Fan2022}
M.~Fan, J.~Wei, W.~Jin, Z.~Xu, W.~Wei, and T.~Liu, ``One step further:
  evaluating interpreters using metamorphic testing,'' in \emph{ACM
  International Symposium on Software Testing and Analysis ({ISSTA})}, 2022.

\bibitem{samek2016evaluating}
W.~Samek, A.~Binder, G.~Montavon, S.~Lapuschkin, and K.-R. M{\"u}ller,
  ``Evaluating the visualization of what a deep neural network has learned,''
  \emph{IEEE transactions on neural networks and learning systems}, 2016.

\bibitem{pace1997}
R.~K. Pace and R.~Barry, ``Sparse spatial autoregressions,'' \emph{Statistics
  \& Probability Letters}, 1997.

\bibitem{cortez2009}
P.~Cortez, A.~Cerdeira, F.~Almeida, T.~Matos, and J.~Reis, ``Modeling wine
  preferences by data mining from physicochemical properties,'' \emph{Decision
  Support Systems}, 2009.

\end{thebibliography}

\end{document}